\documentclass[12pt]{iopart}
\usepackage[normalem]{ulem} 
\usepackage{graphicx}
\usepackage{capt-of}
\usepackage{amssymb}
\usepackage{lineno}
\usepackage{booktabs}
\usepackage[justification=centering,font=footnotesize]{caption}
\usepackage{subfig}
\usepackage[section]{placeins}	
\usepackage{siunitx}
\usepackage{url}
\newcommand{\acs}{ACS}
\newcommand{\apd}{APD}
\newcommand{\tdc}{TDC}
\newcommand{\fpd}{FPD}
\newcommand{\acm}{ACM}
\newcommand{\acsSig}{\texttt{ACS}}
\newcommand{\fpdSig}{\texttt{FPD}}
\newcommand{\fid}{\texttt{FID}}
\newcommand{\lun}{\texttt{LUN}}
\newcommand{\startSig}{\texttt{START}}
\newcommand{\stopSig}{\texttt{STOP}}

\newcommand{\caltdc}{\texttt{CALTDC}}

\newcommand{\overall}{\textit{overall}}

\begin{document}

\title[APOLLO: data reduction and calibration]{Fifteen years of millimeter accuracy lunar laser ranging with APOLLO: data reduction and calibration}

\author{N.R.~Colmenares$^{1,2}$\footnote{ORCIDs: 0009-0008-6736-557X (N.R.~Colmenares); 0000-0003-1236-1228 (J.B.R.~Battat); 0009-0008-0789-2052 (D.P.~Gonzales); 0000-0003-1591-6647 (T.W.~Murphy, Jr.); 0000-0002-8780-8226 (S.~Sabhlok)}, J.B.R.~Battat$^3$, D.P.~Gonzales$^1$, T.W.~Murphy, Jr.$^1$, S.~Sabhlok$^1$}
\address{$^1$ Center for Astrophysics and Space Sciences, University of California, San Diego, 9500 Gilman Drive, La Jolla, CA 92093-0424, USA}
\address{$^2$ Geodesy and Geophysics Lab, NASA Goddard Space Flight Center, 8800 Greenbelt Rd., Greenbelt, MD, 20771, USA}
\address{$^3$ Department of Physics, Wellesley College, 106 Central St, Wellesley, MA 02481, USA}
\eads{\mailto{nicholas.r.colmenares@nasa.gov}}

\maketitle

\begin{abstract}
	The Apache Point Lunar Laser-ranging Operation (APOLLO) has been collecting lunar range measurements for $15$\,years at millimeter accuracy. The median nightly range uncertainty since 2006 is $1.7$\,mm. A recently added Absolute Calibration System (\acs{}), providing an independent assessment of APOLLO system accuracy and the capability to correct lunar range data, revealed a $\sim 0.4\%$ systematic error in the calibration of one piece of hardware that has been present for the entire history of APOLLO. Application of \acs{}-based timing corrections suggests systematic errors are reduced to $< 1$\,mm, such that overall data accuracy and precision are both $\sim 1$\,mm. This paper describes the processing of APOLLO/\acs{} data that converts photon-by-photon range meaurements into the aggregated normal points that are used for physics analyses. Additionally we present methodologies to estimate timing corrections for range data lacking contemporaneous \acs{} photons, including range data collected prior to installation of the \acs{}. We also provide access to the full $15$-year archive of APOLLO normal points (2006-04-06 to 2020-12-27).
\end{abstract}

\section{Introduction}
Modern physics is built upon two primary pillars: general relativity and quantum mechanics, which describe nature at its largest and smallest scales, respectively. The standard model of particle physics, which describes three fundamental forces to a very high accuracy, omits gravity entirely. Of the two pillars comprising modern physics, gravity is less accurately tested than Standard Model forces. Experimental gravitational constraints are challenging to study due to gravity being about $39$ orders-of-magnitude weaker than quantum mechanics, when comparing their scaled coupling constants \cite{PDG_fundamental_forces}.
Very massive objects are necessary for making meaningful measurements of certain aspects of gravitation, and as such, solar system objects have traditionally provided the most practical means of measurement to accomplish this. In our local region, the Earth-Moon system provides a convenient laboratory for probing gravity using lunar laser ranging (LLR). Many leading constraints on gravity (such as the strong equivalence principle\footnote{Recently, improved constraints on the strong equivalence principle were obtained from NASA's MESSENGER mission data~\cite{MESSENGER}, and from a hierarchical stellar triple system consisting of a white dwarf and millisecond pulsar orbiting another white dwarf~\cite{SEP_triple_star}.}) have historically been tested in this natural laboratory, and the goal of modern-day LLR experiments is to improve upon these constraints \cite{murphyLLRReview2013,Nordtvedt:1968zz}.

LLR began in 1969 after the installation of a retroreflector array (composed of fused silica corner cube reflectors, or CCRs) on the Moon by the Apollo 11 astronauts. Subsequent retroreflector arrays were placed on the Moon by Apollo missions 14 and 15 \cite{Chang1972}, as well as by remotely landed Soviet rovers Lunokhod 1 and 2 \cite{Fournet1972}. In LLR, a pulsed laser is fired from the Earth to one reflector array at a time, and the round trip time (RTT) of photon flight is recorded. Each nanosecond of RTT corresponds to $15\,$cm in one-way range; routine range precision and accuracy on the millimeter scale is achieved for the Apache Point Observatory Lunar Laser-ranging Operation (APOLLO) \cite{murphyAPOLLO2008,murphyLLRReview2013,ACS2017,nickThesis}. Observations throughout the course of the Moon's trajectory enable determination of the shape of the lunar orbit, allowing for judgement of the validity of opposing gravitational theories. For example, theories must account for $\sim 10$\,m post-Newtonian (relativistic) effects on the Earth-Moon separation, as measured in the solar system barycenter (SSB) frame \cite{murphyLLRReview2013}. Gravitational tests aside, LLR is additionally sensitive to Earth orientation parameters~\cite{Muller_2014}, secular evolution of the Earth-Moon distance~\cite{Williams_2016}, and the physical properties of the Moon~\cite{Williams_2015,Williams_2001}. A more detailed overview of LLR science deliverables can be found in Ref.~\cite{murphyLLRReview2013}. 
 
APOLLO currently produces the highest-precision LLR measurements. The dominant source of temporal uncertainty for modern LLR stations such as APOLLO is due to reflector geometry being a flat tilted array of many small-aperture corner cubes. Several research groups are investigating next-generation single-element large-aperture fused silica corner cube reflectors (\cite{Currie_2013,Ciocci_2017,Turyshev_2013,He_2018,Araki_2016}) as well as novel bonding techniques for their high tensile strength and low coefficients of thermal expansion \cite{Preston_Merkowitz_2013,Preston_Merkowitz_2014,He_2018}. These next-generation corner cubes will largely address the geometry-induced uncertainty of the original reflector arrays, offering potentially significant improvements to range precision for an equivalent number of collected range photons. 

APOLLO fires powerful laser pulses ($\sim 115$\,mJ, $\sim 90$\,ps FWHM at 20\,Hz, 532\,nm) to retroreflector arrays on the Moon. A small portion of the outgoing light is intercepted by a local retroreflector attached to the secondary mirror of the 3.5\,m telescope at the Apache Point Observatory (APO). The relative timing of the ``fiducial'' (\fid{}) photons from the local corner cube and the returning ``lunar'' (\lun{}) photons yield a differential measurement, effectively determining the separation of the local and remote corner cubes using the same detector and timing electronics. The construction of a differential measurement is important to circumvent the difficulty in making a precise absolute timing measurement due primarily to the temperature dependence of the timing electronics. After each laser pulse (``shot'') the $4\times4$ avalanche photodiode (\apd{}) detector array is turned on (in associated detector activation events called ``gates'') once for the fiducial returns and once for the lunar returns, typically detecting $\lesssim 1$ lunar photon across the array. A custom Computer-Aided Measurement And Control (CAMAC) device called the APOLLO Command Module (\acm{}) is the primary state machine for APOLLO, scheduling gates using a series of counters, registers and comparators referenced against a 50\,MHz clock train. A more detailed description of APOLLO instrumentation, operation and sequence of events can be found in Ref.~\cite{murphyAPOLLO2008}.

More recently, an Absolute Calibration System (\acs{}) was added to APOLLO, to independently assess system accuracy and check for systematic errors \cite{ACS2017}. The \acs{} utilizes a low-power semiconductor saturable absorber mirror (SESAM) fiber cavity laser (different from the main ``range'' laser) whose cavity length is thermally modulated by a phase locked loop (PLL) referenced to a cesium (Cs) time standard in order to stabilize the pulse repetition period. The laser generates $1064$\,nm light pulses of $10$\,ps duration at $80$\,MHz. Individual pulses are picked out of the pulse train to be delivered to the \apd{} during the \fid{} and \lun{} gates resulting in calibration photons overlaid onto range measurements that form a ``ruler'' of optical ``tick marks.'' We can compare what APOLLO determines the timing separation of these tick marks to be against the known truth (a multiple of the 80\,MHz pulse separation of 12.5\,ns), given the exceptional regularity with which \acsSig{} pulses are delivered out of the laser. To distinguish range photons (non-\acsSig{}) from \acsSig{} photons, photon timestamps are compared against predicted range photon return times as well as expected \acsSig{} pulse timings to result in strongly peaked signals of range returns and \acs{} detections respectively. Details of the \acs{} apparatus and performance can be found in Ref~\cite{ACS2017}.

Raw APOLLO data products, or ``runs,'' are the result of photon collection periods spanning $3\mathrm{-}10$ minutes ($3000\mathrm{-}10000$ laser shots) on a single lunar reflector. A run typically contains a few hundred lunar range photons, and a single observing session spanning $1\mathrm{-}1.5$ hours typically contains $5\mathrm{-}10$ runs; further characterization of APOLLO operation can be found in Ref.~\cite{APOLLO_NP_paper_2023}. During each run we record information about photon and gate timing, absolute timestamps provided by a GPS-disciplined quartz oscillator timing standard (XL-DC) and relevant system state parameters. APOLLO's precision timer is a 12-bit 16-channel (one for each \apd{} element) time-to-digital converter (\tdc{}), configured to have $25$\,ps resolution. The data acquired during a run is processed into a final product called a ``normal point.'' A normal point provides a representative measurement of the round-trip-travel time from telescope to reflector and back at some specific epoch, and is the fundamental observable used in subsequent LLR-based analyses \cite{Mueller2019}. 

Utilization of the \acs{} allowed us to uncover a $\sim 0.4\%$ systematic error in the \tdc{} calibration, which impacts APOLLO measurements at the millimeter scale; fortunately, we are able to apply timing corrections to account for this effect to even the earliest of APOLLO data. The rest of this paper describes how APOLLO's calibrated normal points are generated. A full description of APOLLO's performance using normal point statistics is presented in a companion paper~\cite{APOLLO_NP_paper_2023}.

\section{Packaging of Range Photon Timing}
\label{sec:photon_packaging}
To motivate the reduction procedure and application of \acsSig{} corrections, we first summarize the sequence of APOLLO events and how a range measurement is formed. It is critical to know to a high degree of precision when the range laser has fired, as scheduling of gates is referenced against this time. However, the firing time of the range laser cannot be controlled to enough precision given its large jitter ($1.6\,\mu s$). A timing ``anchor'' is used to obtain a low-jitter ($\sim 20$\,ps) measurement in the form of a fast photodiode (\fpd{}), the signal from which is present for every laser shot. In contrast, the \apd{} operates in single-photon mode, with $\sim 100$\,ps of jitter per photon, and may not have detections for every laser shot. An unfortunate consequence of utilizing the \fpd{} as a timing anchor is that it is not a differential measurement with respect to the returning lunar range photons. The \fpd{} signal is handled on one reserved channel of the \tdc{} that will have a different temperature dependence (and different electronic and optical paths to arrive there) compared to the other 15 \tdc{} channels. The differences between \fpd{} and \lun{} signal handling would cause the timing between the two to drift over time; as such, we need to tie our timing ``anchor'' to the lunar photons on a run by run basis by relating the two to the fiducial photons, which take the same electronic and optical paths as the lunar photons. Relating the \fpd{} and \fid{} signals for each run is a relatively simple task given the presence of one or more photon events in most \fid{} gates, providing ample statistics.

\begin{figure}[tbh]
        \centering
        \includegraphics[width=1.0\textwidth]{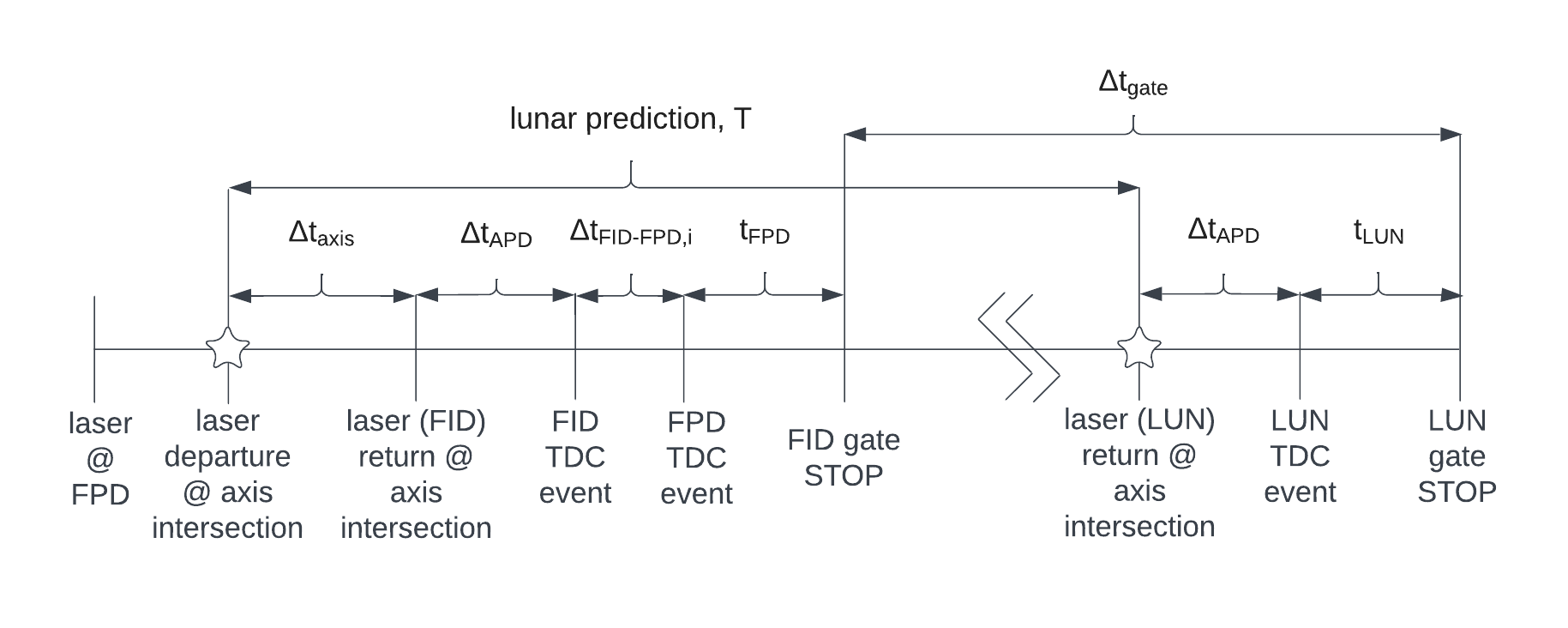}
        \caption{\label{fig:events_timeline}
	Rough sequence of notable APOLLO events for a particular laser shot fired with one fiducial and one lunar photon returned in their respective gate types, in the same detector channel. $\Delta t_{\mathrm{FID-FPD,i}}$ indicates an individual event that contributes to the aggregate fiducial signal $\Delta t_{\mathrm{FID-FPD}}$ seen in equation~\ref{RTT_equation} and discussed in the text. $\Delta t_{\mathrm{APD}}$ denotes the time between laser light returning through the telescope axis intersection and receipt of a photon event at the \tdc{}; note this term cancels out in equation~\ref{RTT_equation}.
        }
\end{figure}

Figure~\ref{fig:events_timeline} shows a crude timeline of APOLLO events for an idealized scenario of a single laser shot in which a fiducial photon and a corresponding lunar gate return photon are present in the same detector channel. The range laser fires, generating a \fpd{} event a short time later as range laser photons leave the APOLLO apparatus. The telescope axis intersection is chosen as the fixed reference point to measure time between the outgoing and returning signal. This reference point is insensitive to telescope motion, and we know to a high degree of accuracy how long it takes light to travel along the optical axis between the axis intersection and the fiducial corner cube based on measurements using a theodolite. 

The \fpd{} signal (henceforth distinguished from the \fpd{} itself by a different font: \fpdSig{}) alerts the \acm{} that the laser has fired, thereby initiating the \fid{} gate and scheduling the corresponding \lun{} gate about 2.5 seconds in the future. A cable-delayed version of the \fpdSig{} is also fed to the \tdc{} within the fiducial gate. A time $T$ (the prediction time, \emph{if} our prediction were perfect) after the transmitted laser pulse reaches the telescope axis intersection, range photons again cross the same reference point after returning from the Moon. Shortly thereafter the light reaches the \apd{} which produces a \startSig{} signal that is sent to the \tdc{}.

Although the fiducial and lunar photon return times are asynchronous relative to the 50\,MHz clock signal used to schedule APOLLO events, we are able to predict when those returns should arrive (based on the \fpdSig{}), and schedule gates to the nearest clock pulse such that range returns uniformly populate a 20\,ns interval in \tdc{}-space, colloquially referred to as a ``slug,'' for both gate types. For each \fid{} and \lun{} gate, we record the \tdc{} reading for each channel that triggered, as well as the last clock pulse in the detection gate using a counter that tracks the number of clock pulses elapsed. Fiducial gate records contain information on when the \fpd{} fired, and lunar gate records contain the predicted \tdc{} timestamp for lunar photons based on the \fpd{} timing from the associated fiducial gate. 

During the reduction procedure, \tdc{} calibrations obtained from a process called ``\caltdc{}'' --- executed before and after each run --- are applied to the raw data. During \caltdc{}, the \tdc{} sends common \startSig{} and \stopSig{} signals (derived from the 50\,MHz clock train) to all channels 1000 times at 1\,kHz each for 5 different \startSig{}--\stopSig{} intervals in integer multiples of 20.00\,ns. A quadratic function is fit to each channel's results to provide a conversion between TDC number and an actual timestamp for any photon event---more information on this process can be found in Ref.~\cite{murphyAPOLLO2008}. Various data integrity checks are also performed, such as checking that counters are not missing any clock pulses. Below is a simplified description of the remaining reduction procedure, followed by a detailed description of each step.

\begin{enumerate}
	\item Subtract the \fpdSig{} timestamp from each \fid{} event timestamp to produce a relative timing signal for each channel of the \apd{}. Combine these into a single ``aggregate'' fiducial signal, thereby establishing channel-to-channel timing corrections as well as characterizing range laser and detector timing performance. \label{itemone}
	\item Fit a physically motivated functional form to the ``aggregate'' fiducial signal, which is later used to inform the lunar functional fit.
	\item For each lunar event timestamp, form lunar ``residuals'' by applying the channel offset corrections determined in step~\ref{itemone} to each lunar event and subtracting the prediction \tdc{} value for that laser shot.
	\item Remove any linear temporal trend in the lunar residuals across the run duration and form a histogram of the result.
	\item Perform a one-parameter sliding fit between the lunar and fiducial histograms and record the result. The shape, size and background vertical offset are pre-determined from the \fid{} signal treatment convolved with a trapezoid representing the libration-specific apparent reflector profile.
	\item Add back in the linear trend and prediction value at a calculated reference time.
\end{enumerate}

		Recall that most shots will not have a photon event in both fiducial and lunar gates for the same \apd{} channel. To utilize all lunar events for our final data product, we form an aggregate fiducial signal to act as a timing reference for every lunar event.
This aggregate fiducial signal is constructed by comparing the arrival times of each \fid{} photon against the laser fire time (the \fpdSig{}). The difference in \fid{} and \fpdSig{} arrival times is effectively constant over the duration of a run, but slightly different for each channel (due to differences in electronic delays of each). Relative channel corrections $\delta_{\mathrm{ch}}$ are determined on a per-run basis by correlating each channel against a reference signal (arbitrarily chosen to be the first \tdc{} channel).
	We can then apply each channel's timing correction and combine data from all channels to produce a strongly peaked signal (the ``aggregate''). The width of the aggregate fiducial is set by the laser pulse width and detector/timing electronics jitter. It has an asymmetric tail due to photoelectron diffusion into the depletion region of the \apd{} \cite{murphyLLRReview2013}. A function (gaussian with asymmetric tail) is then fit to the aggregate signal. From here, the RTT for an individual lunar photon is calculated from: 
	\begin{equation}
		\label{RTT_equation}
		RTT = \Delta t_{\mathrm{gate}} + t_{\mathrm{FPD}} + \Delta t_{\mathrm{FID-FPD}} - t_{\mathrm{LUN}} - \delta_{\mathrm{ch}} + \Delta t_{\mathrm{axis}},
	\end{equation}
	where:
	\begin{itemize}
		\item	$\Delta t_{\mathrm{gate}}$ is the number of clock oscillator periods between the ends of the \fid{} and \lun{} gates for this laser shot (a multiple of 20.00\,ns),
		\item	$t_{\mathrm{FPD}}$ is the time of the \fpdSig{} signal relative to the end of the \fid{} gate,
		\item	$\Delta t_{\mathrm{FID-FPD}}$ is the fitted aggregate fiducial timing,
		\item	$t_{\mathrm{LUN}}$ is the lunar photon return time relative to the end of the lunar gate,
		\item	$\delta_{\mathrm{ch}}$ is the relative channel correction,
		\item	$\Delta t_{\mathrm{axis}}$ is the round trip light travel time from the axis intersection to the local CCR and back.
	\end{itemize}
	
\begin{figure}[tbh]
	\centering
	\includegraphics[width=0.85\textwidth]{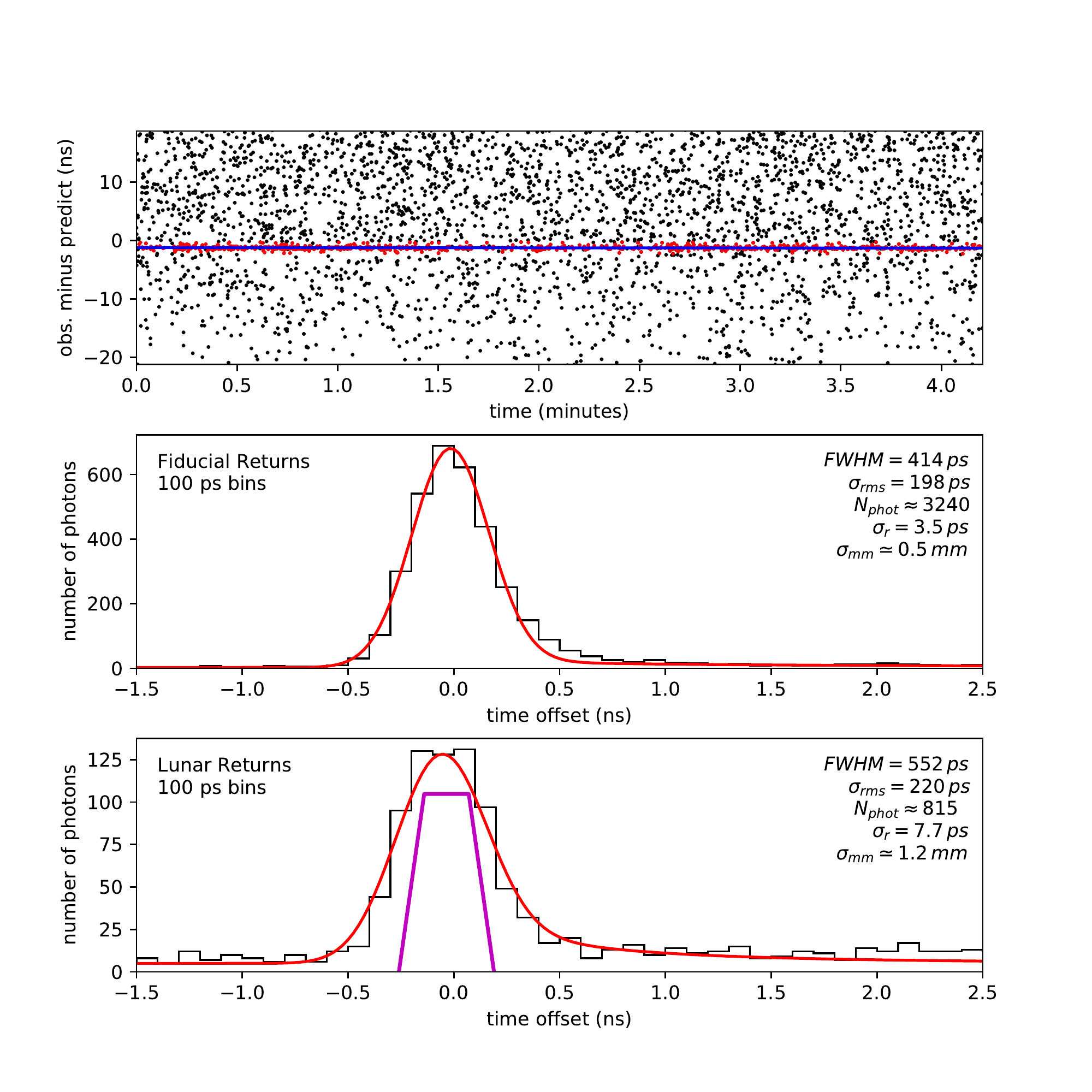}
	\caption{\label{fig:190416k_resids} 
	Top: lunar gate photon timing residuals versus time within run, for a run on the night of 2019 April 16, after subtracting the predicted round-trip time for each shot. The linear residual is represented by a blue line, and photons within some window surrounding the fit are colored red, indicating they are candidate lunar range photons. Middle: the aggregate fiducial signal with its fitted centroid subtracted. The asymmetric fit function is overlaid in red. Bottom: histogram of the difference between lunar events and the fitted blue line in the upper plot, with the retroreflector array timing profile for this observation shown in magenta. The red line is the fiducial fit convolved with the reflector profile and shifted horizontally to optimally align with the lunar histogram.}
\end{figure}

Lunar photon timing residuals are obtained by subtracting the predicted return time on a shot-by-shot basis from the calculated RTTs (with channel corrections applied), resulting in a strongly peaked signal similar to the aggregate fiducial, as seen in Figure~\ref{fig:190416k_resids}. A line is fit to the residuals vs. time within the run to capture the offset and slope. Lunar returns that fall within a windowed region of the fitted line are tagged as ``registered'' lunar returns (red points in Figure~\ref{fig:190416k_resids}, top), and a refined linear fit is performed on these registered photons. A representative time is formed for the run by finding the mean of registered lunar return times. The fitted line is subtracted from the residual signal to result in a new residual histogram. A one parameter fit is performed on the latest residual histogram by sliding the aggregate fiducial signal fit, convolved with the reflector tip/tilt profile at the time of ranging, until best-overlaid. During the fit we impose a constraint that the area under the fit function be matched to the area under a windowed region of the lunar histogram and an additive constant is set to match the measured background. The representative time of the registered lunar photons is used to calculate a predicted round-trip time for the run \emph{as a whole}, and is combined with the centroid of the lunar histogram fit, plus the representative timing offset of the fitted trend line; the ensuing number is the fundamental lunar range measurement: a representative round-trip time for the \emph{entire run}---the aforementioned ``normal point.''

\subsection{Clock corrections}
\label{subsec:clock}
Prior to installation of the \acs{} in 2016 August, APOLLO only used the GPS-disciplined Agilent XL-DC clock as its timebase. After the \acs{} was installed, APOLLO utilized either the newer Symmetricom cesium clock or the XL-DC clock as its timebase. Since the arrival of the cesium clock in 2016 February, a Universal Counter (UC) has monitored the relative frequency and phase differences between the two timebases. A methodology for determining corrections to round trip times informed by the XL-DC clock was developed based on the UC comparison data~\cite{APOLLO_clock2017}. 

Runs that use the XL-DC as APOLLO's timebase without corresponding UC comparison data are correctable in the manner described in~\cite{APOLLO_clock2017}, which gives a single RTT adjustment and uncertainty modification per normal point; this encompasses runs before 2016 February 12 and runs from 2017 December 21 to 2018 May 11, when the cesium clock underwent repairs. However, runs that use the XL-DC as APOLLO's timebase but \emph{do} have corresponding UC comparison data are eligible to receive slightly more direct clock corrections. The concept is effectively the same as in~\cite{APOLLO_clock2017}, but since we have UC comparison data every 10 seconds, we may apply clock corrections determined for each 10 second interval (using comparison data averaged over the closest three to four 10 second intervals, for smoothness) to individual lunar range photon round trip times instead of a singular correction to the normal point's final representative round trip time and uncertainty. Runs from 2016 February 12 to 2017 January 4 and from 2017 February 3 to 2017 February 15 (during which we reverted to using the XL-DC while investigating an issue related to the cesium clock) are correctable using this slightly modified version of~\cite{APOLLO_clock2017}'s methodology.

\subsection{``Long-term'' channel corrections}
\label{subsec:longterm}
Raw APOLLO normal point uncertainties are based only on data from individual runs, but there may still exist systematic trends from RTT differences between channels. By comparing individual-channel RTTs against the weighted mean of RTTs over all channels for a given run, it is possible to assess how consistent and compatible a particular channel's timing is relative to the overall collection. A single run will show some scatter among individual-channel RTT calculations that may or may not be statistically aligned to the whole, which becomes easier to discern for runs that acquire a large number of photons.

One run alone is not enough to elucidate systematic trends, so in practice we have analyzed these patterns over long timespans corresponding to distinct hardware periods. We restrict the study to include runs that accumulated at least $500$\,photons so that individual channels would have enough signal to produce useful statistics. Over the period of time represented in this analysis, we could assess for each channel whether any systematic offset existed, and if the excursions from the channel's weighted mean exceed that expected from estimated uncertainties. In other words, a chi-squared measure of departures from the channel weighted mean indicates whether the scatter is consistent with or exceeds estimated uncertainties.

We may then use the results of these ``long-term'' analyses to apply corrections to individual channels equivalent to each channel's deviation from the weighted mean of the whole; this has the effect of tightening-up the distributions of range returns such that the functional fit mentioned in the previous section performs better. Additionally, we can account for the disconnect between expected and observed scatter by applying a measure of excess spread to the estimated normal point uncertainties. These error inflations are computed in two ways: as a multiplicative scaling of the original errors, or (more appropriately, we think) as a root-sum-square (RSS) error term to be added in quadrature to the raw normal point uncertainty. Note however that these error inflations are provided to analysts as suggestions; APOLLO normal point uncertainties have not been adjusted by these suggested amounts. Recommended error inflations are discussed in more detail in~\cite{APOLLO_NP_paper_2023}.

\section{The \acs{} signal}
\label{sec:acs_signal}
A detailed description of the \acs{} hardware and integration into the rest of the APOLLO apparatus is provided in Ref.~\cite{ACS2017}. Here, we describe the salient characteristics necessary to understand the \acsSig{}-based timing calibration of APOLLO data.

The \acs{} laser repetition rate is 80\,MHz, locked to a stable cesium frequency standard. \acsSig{} photons are coupled into the same \apd{} detector array and timing system that is used for the lunar range measurement. Although \acsSig{} photon pulses are 12.500\,ns apart, the 80\,MHz pulse train has five possible positions relative to the 50\,MHz clock pulse that forms the \stopSig{} signal for the \tdc{}. These five possible positions produce an \acsSig{} ``comb'' in \tdc{}-space, with comb ``teeth'' separated by $n\times2.5$\,ns, where $n$ is an integer.

\begin{figure}[t]
        \centering
        \includegraphics[width=0.9\textwidth]{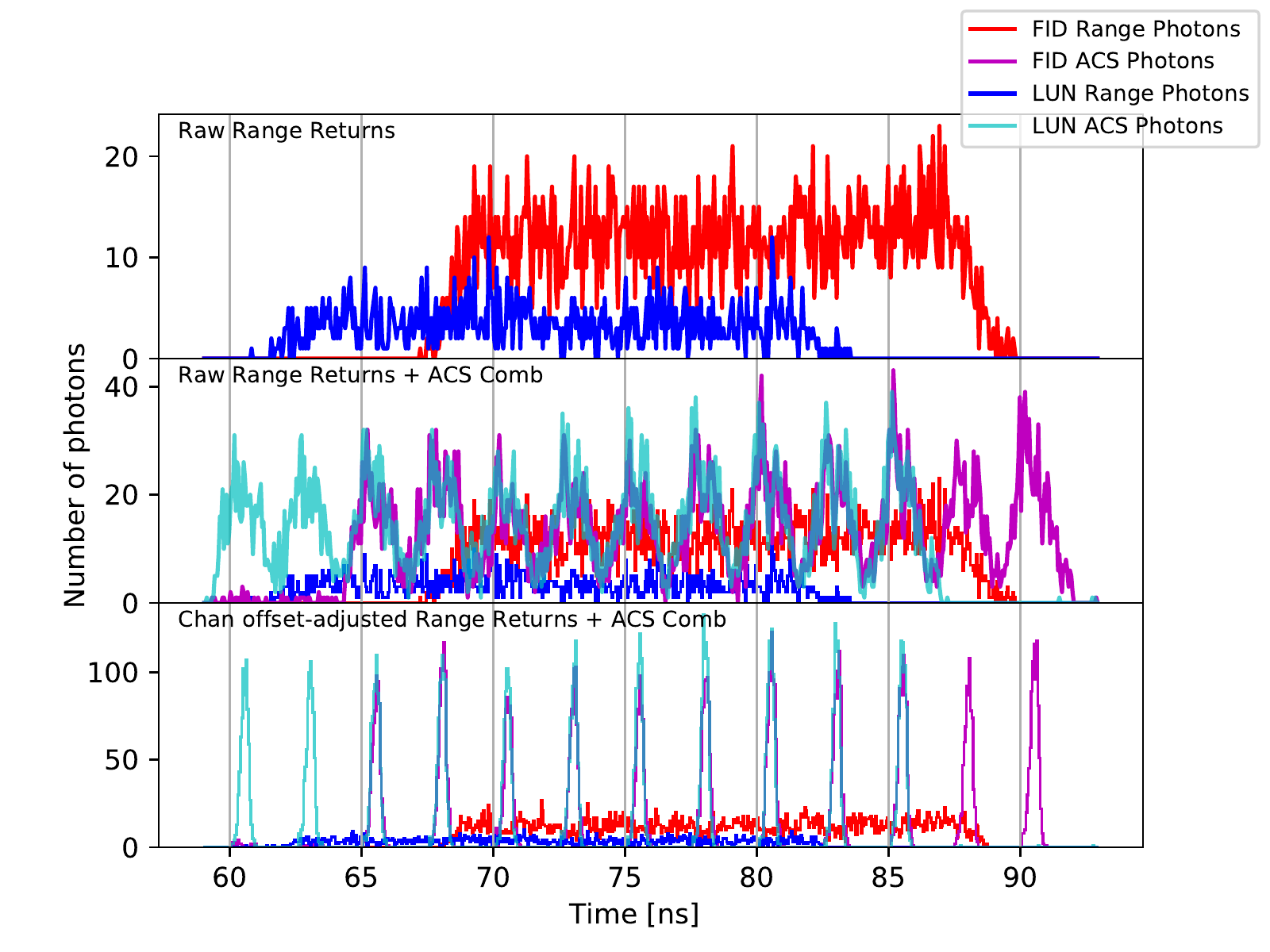}
	\caption{\label{fig:comb_overlay_demonstration} Histograms of time-converted range and \acs{} returns for the same run as in Figure~\ref{fig:190416k_resids}. The top and middle plots do \emph{not} include relative channel corrections, while the bottom \emph{does}.
        }
\end{figure}

Even though the calibration and lunar range (or fiducial) photons may arrive in a single gate of the \apd{}, timing within the gate can be used to tag detected photons as \acsSig{} or \lun{} or \fid{}. An example of this is given in Figure~\ref{fig:comb_overlay_demonstration}. The top panel shows the \fid{} and \lun{} photons (aggregated over all \apd{} channels), while the middle panel includes the \acsSig{} photons. The \acsSig{} peaks are broad because timing corrections between different channels of the \apd{} have not been accounted for. Once those corrections are applied, the \acsSig{} ``comb'' is clearly seen (such as in the lower panel of Figure~\ref{fig:comb_overlay_demonstration}). Comparing the measured locations of the peaks to the known intrinsic separation of $n\times2.5$\,ns provides a calibration of the APOLLO timing system.

The quality of the \acs{} calibration depends on the precision with which the centroid of each tooth is known. That, in turn, depends on the number of photons in the tooth. The \acsSig{} light level is limited to $\sim1.5\,\frac{\mathrm{photons}}{\mathrm{gate}}$ (we avoid higher rates because an \apd{} element that has detected an \acsSig{} photon would be blind to a subsequent lunar/fiducial return signal in that gate). A typical lunar ranging run contains 5000 range laser shots (gates), spread across 15 active \apd{} channels, and an \acsSig{} comb consists of $\sim 10$ teeth. So each tooth of an \acsSig{} comb for a given \apd{} channel contains about 50 of the 7500 \acsSig{} photons.  The \acsSig{} is often disabled during the first $\sim 20\%$ of a run to ensure proper lunar range signal acquisition, bringing the effective \acsSig{} shot duration down from 5000 to $\sim4000$, and the  per-tooth per-channel \acsSig{} photon count down to 40.  A typical comb tooth is $\sigma_\gamma \approx 130$\,ps wide for data prior to August 2019, and $\sigma_\gamma \approx 80$\,ps after (an upgrade to the \apd{} in August 2019 improved the timing precision). The tooth centroid is therefore determined to a precision of $\frac{\sigma_\gamma}{\sqrt{N}} \approx 21$\,ps ($13$\,ps) before (after) the hardware upgrade when $N \approx 40$. Range photon timing corrections rely on the difference of tooth positions, which further degrades the calibration precision by a factor of $\sqrt{2}$ to 30\,ps (18\,ps), or 5\,mm (3\,mm) of one-way range. Given that the typical APOLLO normal point uncertainty is 1--2\,mm, and that the typical scale for \acsSig{}-based timing corrections of range data is 3--7\,mm, we need a more precise method to calibrate APOLLO data.

\begin{figure}[htb]
  \centering
  \includegraphics[width=0.8\textwidth]{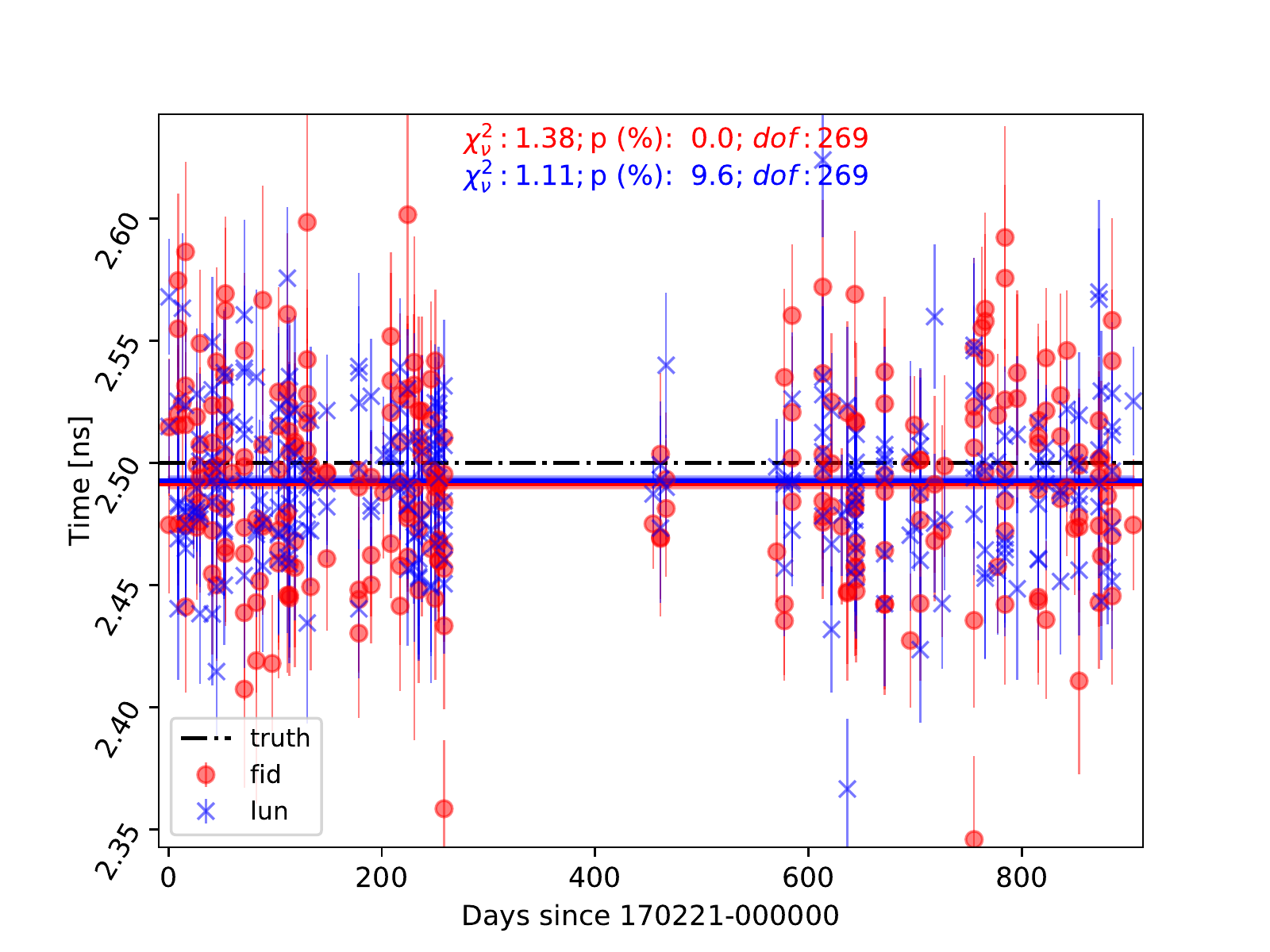}
	\caption{\label{fig:toothSep_plot} Tooth separations between an adjacent pair of teeth, for a single \apd{} channel, over several years. Each gate type (\fid{} or \lun{}) is plotted separately. The $\chi_\nu^2$, corresponding p-values$*100$ and the degrees of freedom (DOF) are shown for each gate, relative to the weighted mean which is displayed as colored solid horizontal lines. The horizontal black dash-dotted line shows the expected $2.500$\,ns tooth separation if APOLLO counted time perfectly. The weighted means for each gate lie approximately $10$\,ps below $2.500$\,ns, or $\sim 0.4$\%, which is common across channels and tooth pairs and indicates an undercounting of time in the \tdc{} -- a feature which will be discussed in subsequent sections of this paper. The two large gaps in the data (days 259 to 454 and 467 to 570) are due to a reference clock failure that halted \acs{} operations, and a scheduled observatory maintenance shutdown, respectively.}
\end{figure}

To enhance the comb tooth centroid determination, one might consider aggregating \acsSig{} data taken across many runs, thereby increasing the number of total \acsSig{} photons in each comb tooth. Unfortunately, temporal drifts in the comb tooth location in \tdc{}-space make this approach unviable.  Instead, we accumulate measurements of the separation \textit{between} comb teeth across runs. As shown in Figure~\ref{fig:toothSep_plot}, these separations are stable over timescales of years. Reduced chi-squared measures suggest the presence of extra structure and/or underestimated uncertainties (as seen for the \fid{} result in Figure~\ref{fig:toothSep_plot}); subsequent investigations revealed no extra underlying structure, so uncertainties were inflated to result in reduced chi-squared measures near unity. As described in Section~\ref{sec:rangeCalibrationWithAcs}, we leverage this stability to arrive at an \acsSig{}-derived timing calibration with an accuracy of $\sim 0.5$\,mm.

It is additionally worth noting that the weighted means of tooth separations for both gates in Figure~\ref{fig:toothSep_plot} lie below the expected separation of $2.500$\,ns by approximately $0.4$\% ($\sim 10$\,ps), demonstrating that the \tdc{} tends to undercount the time interval between this pair of teeth on average. This undercounting of time is a common feature across channels and tooth pairs, providing a clue that APOLLO may systematically count time too slowly in the \tdc{}. This phenomenon will be seen repeatedly in successive sections of this paper.

Any APOLLO run either has sufficient contemporaneous \acsSig{} data or it does not. In the former case, the \acsSig{} data can be used to calibrate the ranging data on a photon-by-photon basis using an approach described in Section~\ref{sec:rangeCalibrationWithAcs}.  If \acsSig{} data is not available or inadequate for a given run, calibration options still exist, as described in Section~\ref{sec:rangeCalibrationWithoutAcs}.

\section{Range timing calibration with contemporaneous \acsSig{} data}
\label{sec:rangeCalibrationWithAcs}

APOLLO runs containing sufficient contemporaneous \acsSig{} information allow us to calibrate individual \fid{} and \lun{} photon timings. Photon timing corrections are determined by identifying the position of a range photon relative to a representative \acsSig{} comb (aggregated over years). This section describes the photon-by-photon calibration procedure in greater detail.

As suggested in section~\ref{sec:acs_signal}, we observe that \acsSig{} combs from one run to the next over timescales of years largely share the same pattern of tooth \emph{separations} but suffer from temporal shifts in \tdc{}-space. Aggregation of tooth separations relative to a chosen reference tooth allow us to form high signal-to-noise representative \acsSig{} combs, or ``template'' combs (one for each channel/gate-type combination), while effectively decoupling the templates from \tdc{}-space. Template combs are later matched to \acsSig{} combs from individual runs and photon-by-photon range timings may then be calibrated based on their position relative to the resulting comb. Occasionally APOLLO experiences hardware changes, which are typically accompanied by the formation of a new set of ACS template combs.

\subsection{The Calibration Method}
\label{subsec:calibrationMethod}
Having defined a template comb for each channel and gate type, we can form a calibration mapping to convert photon \tdc{} records to time, distinct from the \caltdc{}-derived calibration functional form. We assert that the teeth of the template comb should have a fixed pitch of $2.5$\,ns, relative to a chosen reference tooth of the template comb (the same tooth for every channel/gate combination). The choice of reference tooth is kept fixed across hardware periods in order to connect ACS results in a consistent manner. For photon events that fall between template teeth, interpolation is used to permit calibration for any location within the span of the template comb. Any photon events that fall outside the template comb see an extrapolated portion of the calibration mapping, based on the template comb's mean tooth separation.
The final step is to determine, for runs with contemporaneous \acsSig{} data, where each run's \acsSig{} combs \emph{actually are} within \tdc{}-space. We accomplish this by sliding each template comb until best overlaid with the corresponding comb in an individual run (the ``daughter'' comb). The alignment between a given template comb and a corresponding daughter comb is optimized via a one-parameter least squares fit. The resulting fitted comb position tells us where in \tdc{}-space the previously determined calibration mapping should be placed. We refer to the fitted template combs as ``idealized'' combs to avoid confusion, given ``daughter'' combs are formed from contemporaneous \acsSig{} photons in a given run, and ``template'' combs are decoupled from \tdc{}-space.

Now that we have a calibration mapping derived at the location of each template tooth, and a best-fit position for the \emph{actual} comb (the ``idealized'' comb), timing records for range photons can be obtained. Each range photon has a measured position in \tdc{}-space which can be used to compute the corresponding \acsSig{}-derived time calibration based on their proximity to idealized comb teeth. The data reduction process as described in Section~\ref{sec:photon_packaging} continues normally after range photons are time-calibrated.

One consequence of the \acs{} methodology described above that users of the data should be aware of is that normal point results will change slightly over time as \acsSig{} calibrations are improved from aggregation of new \acsSig{} data. \acsSig{} template combs are aggregated within time periods of fixed hardware configurations of APOLLO. Certain hardware changes to APOLLO may influence the structure of the \acsSig{} template combs, so ``new'' template combs are started any time these changes are expected or observed to impact \acs{} results.

\subsection{Range Error Statistics}
\label{subsec:rangeErrStats}
At the end of each run's data processing, we seek to summarize the scale of timing inaccuracies that were corrected by using the \acsSig{}-derived calibration into a single representative differential estimate. We may then study how \acsSig{} corrections distribute themselves over time and/or collections of runs. Since using the \caltdc{} calibration is the default operation in absence of sufficient \acsSig{} information, we define an \acsSig{} timing correction for an individual \fid{} or \lun{} range photon as the difference between its calibrated timestamp as determined by \acsSig{} and its calibrated timestamp as determined by \caltdc{}:
        \begin{equation}
                \label{individual_correction}
                C_{\texttt{i}} \equiv t_{\acsSig{}, \texttt{i}} - t_{\caltdc{}, \texttt{i}}.
        \end{equation}
The mean of timing corrections to \fid{} and \lun{} range photons are obtained separately for those range photons considered to be contributors to the actual signal, and differenced such that
        \begin{equation}
                \label{differential_correction}
                C_{\texttt{differential}} \equiv C_{\fid{}} - C_{\lun{}},
        \end{equation}
where $C_{\fid{}}$ and $C_{\lun{}}$ are the mean \fid{} and \lun{} timing corrections, respectively\footnote{Note: while in Ref~\cite{nickThesis} the discussion centers around timing \emph{offsets} rather than timing \emph{corrections} (one is the negative of the other), the corresponding differential quantity defined at the end of Section 5.3.1 of Ref~\cite{nickThesis} had an erroneous negative sign applied; as such, any references to timing offsets within or after Section 5.3.2 should be interpreted as timing \emph{corrections} instead.}.

\begin{figure}[tbh]
        \centering
	\includegraphics[width=0.9\textwidth]{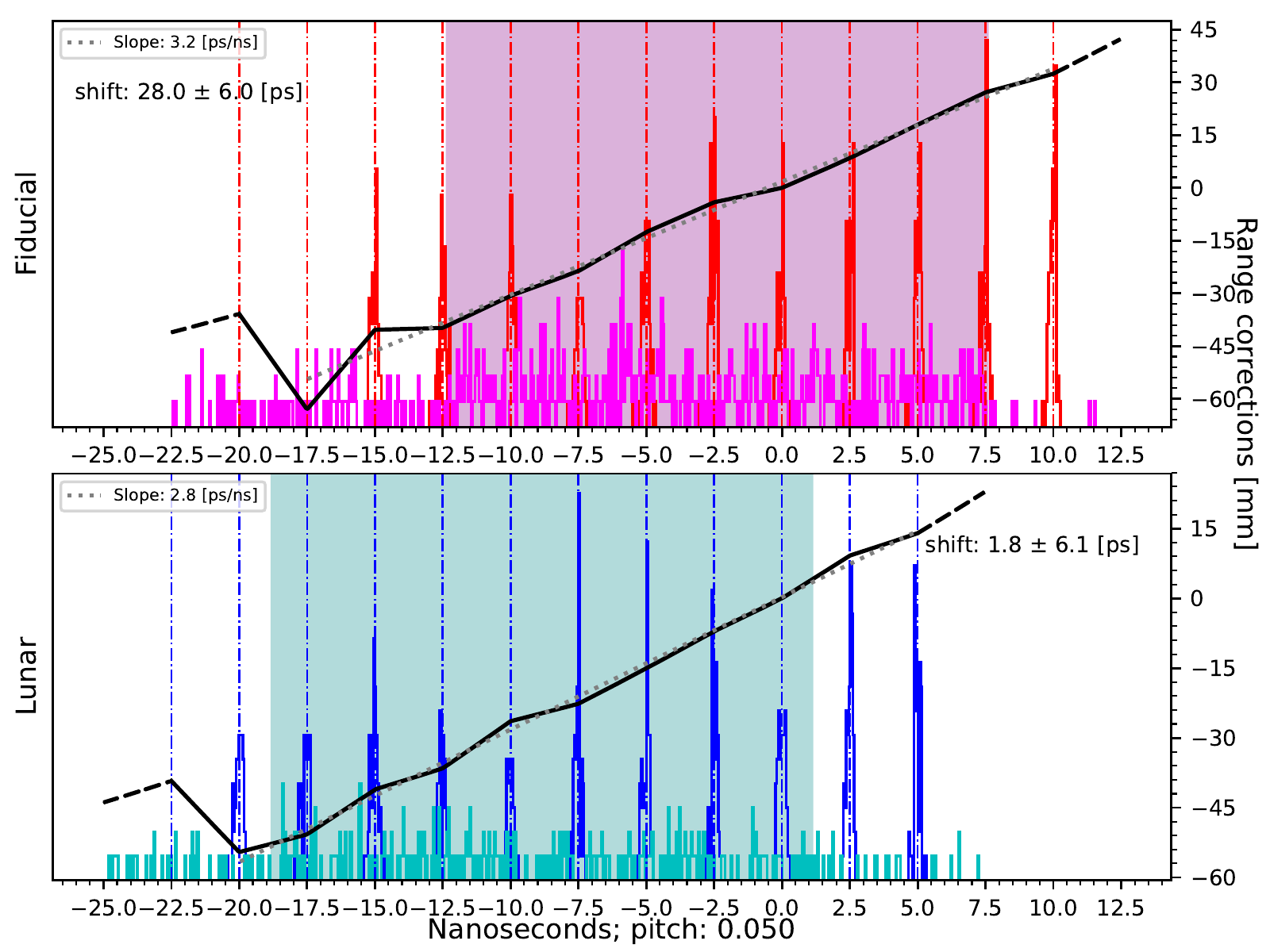}
	\caption{\label{fig:real_mapping} Actual data for the same run pictured in Figure~\ref{fig:190416k_resids} for the same channel as in Figure~\ref{fig:toothSep_plot}. Vertical dash-dot lines denote positions of template comb teeth which have been shifted to best align to the daughter comb shown as the red and blue spikes. This shifted comb is referred to as the ``idealized'' comb in the main text. Teeth on the ends of the template comb are sparsely populated, such that noise effects are more obvious. The dark sloping line indicates interpolated timing corrections, derived from the template comb as defined in Equation~\ref{individual_correction}. Extrapolated portions of the timing corrections are indicated with dashed black lines. The timing corrections, whose scale in picoseconds appears on the right, demonstrate a mostly smooth structure that adds confidence to the interpolation approach. Dotted gray lines denote a linear fit to the timing corrections, excluding the first template comb tooth in each gate. The fitted slopes for each gate are $\sim 3\,\frac{\mathrm{ps}}{\mathrm{ns}}$, or $0.3$\% --- another preview of the eventual $0.4$\% systematic offset from truth discovered when averaged over all channels. The histograms shown in magenta and cyan are the raw \fid{} and \lun{} range returns, respectively. Lightly shaded regions appearing in magenta and cyan atop the histograms denote the $20$\,ns span that registered range photons land within for the \fid{} and \lun{} gates respectively; notice the $20$\,ns range return region is well-covered by the \acsSig{} comb for each gate type. 
        }
\end{figure}

We invite the reader to refer to Figure~\ref{fig:real_mapping}, which displays an example result of the process described earlier in this section, for an illustrative example. More template/idealized comb data points appear than the number of daughter comb teeth, due to slightly different regions of \tdc{}-space being sampled by \acsSig{} photons over time as we learned how to properly utilize the \acs{} system. The slopes of interpolated portions of the correction mapping indicate how template/idealized \emph{adjacent} teeth are spaced --- a positive slope indicates the teeth are too close, while a negative slope indicates teeth are too far apart. Additionally, it is worth noticing the significance of the sign of corrections themselves, relative to what side of the reference tooth (where $x=0$ in Figure~\ref{fig:real_mapping}) they fall on. Negative corrections on the left side of the reference tooth and positive on the right side of the reference tooth both correspond to comb teeth being separated less than expected, relative to the reference tooth, representing compressed time. Figure~\ref{fig:real_mapping} suggests that, relative to a chosen reference tooth, all other teeth in this template comb are systematically spaced too closely compared to truth --- a tendency that was first hinted at in Figure~\ref{fig:toothSep_plot} due to the weighted means being systematicaly lower than $2.500$\,ns. 

The gray dotted lines in Figure~\ref{fig:real_mapping} indicate this channel counts time too slowly in the \tdc{} part of the timing system --- $\sim 60$\,ps less than truth, over a $20$\,ns window in each gate type, as suggested by the fitted slopes. Similar to Figure~\ref{fig:toothSep_plot}'s weighted means being offset from $2.500$\,ns, the slope assessment for this particular channel provides another preview of what becomes a newly discovered overall $0.4$\% systematic inaccuracy in the \tdc{} when analyzed for all channels. 

\begin{figure}
        \centering
        \includegraphics[width=0.8\textwidth]{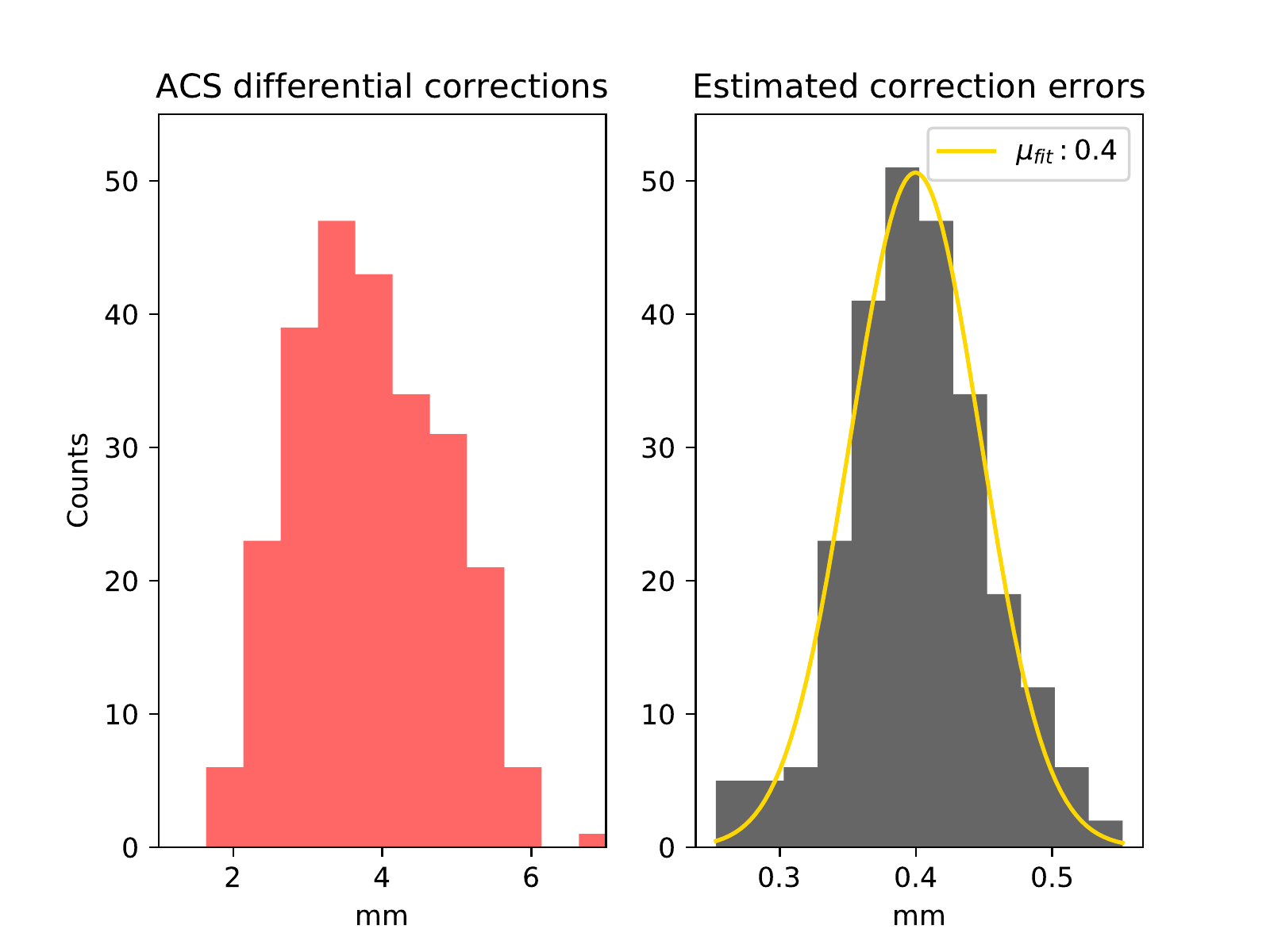}
	\caption{\label{fig:acs_run_correction_compare} Left: distribution of \acsSig{} corrections ($C_{\texttt{differential}}$) for runs using the cesium clock as the APOLLO timebase in the initial \acsSig{} data collection period, corresponding to 2017 February 21 through 2019 August 14. The range of \acsSig{} corrections is reflective of variations in the prediction used to acquire a range signal -- a well-understood phenomenon that is expanded on in Section~\ref{sec:rangeCalibrationWithoutAcs}. Right: distribution of estimated \acsSig{} correction errors for the same dataset. A best-fit gaussian is overlaid, with a centroid of $0.4$\,mm.}
\end{figure}

Figure~\ref{fig:acs_run_correction_compare} shows the distribution of \acsSig{}-based corrections and their uncertainties for APOLLO data that used the cesium clock between 2017 February 21 and 2019 August 14. The left histogram in Figure~\ref{fig:acs_run_correction_compare} suggests that APOLLO has a systematic tendency to underestimate the one-way range by $\sim 4$\,mm, resulting from the \tdc{}'s systematic undercounting of time. The right-side histogram indicates sub-mm accuracy of \acsSig{} corrections, and thus of APOLLO data once corrected. A more detailed discussion of this \tdc{} systematic error effect and how it can be utilized to predict \acsSig{}-based timing corrections will be presented in Section~\ref{sec:rangeCalibrationWithoutAcs}.

Occasionally APOLLO experiences hardware and/or configuration changes which have implications regarding range accuracy. In terms of \acs{} operation and usage, these hardware changes are usually accompanied by the formation of a new set of \acsSig{} template combs and systematic error statistics. More information regarding changes to APOLLO hardware and the system clock can be found in Ref~\cite{APOLLO_NP_paper_2023}.

\section{Range timing calibration without contemporaneous \acsSig{} data}
\label{sec:rangeCalibrationWithoutAcs}
When the lunar range signal is weak, delivery of \acs{} light onto the \apd{} is deliberately disabled. The addition of \acsSig{} photons interferes with telescope pointing feedback derived from \apd{} event rates, and can effectively mask the range signal underneath a ``floor'' of detections when coupled with difficult observing conditions. Despite these runs lacking concomitant \acsSig{} data, we still wish to calibrate their lunar range measurements based on aggregate \acsSig{} information. Here we present a technique to utilize \acsSig{} correction trends to determine trustworthy corrections on a run-by-run basis for those runs without concurrent \acsSig{} photons, which we refer to as ``non-\acs{}'' runs.

Lacking \acsSig{} information for a run, a straightforward substitute correction could be obtained from the centroid of the distribution shown in Figure~\ref{fig:acs_run_correction_compare}, which we label as the ``\overall{}'' distribution. We observe, however, that we may be able to effectively predict a run's \acsSig{} correction by having noticed a tight correlation between derived \acsSig{} corrections and the positioning of \lun{} and \fid{} photons in \tdc{}-space. Imagine for a moment that the cyan histogram (and correspondingly the cyan shaded region) in the lower subplot in Figure~\ref{fig:real_mapping} moves to the right while everything else in the entire figure remains fixed. The average value of \acsSig{} corrections in the lunar gate for this channel would therefore be different (more positive). Now imagine this were the case for every channel in this run. Consequently, the \acsSig{} correction for the entire run ($C_{\texttt{differential}}$) would be different.

In reality, the range photon distributions do move around in \tdc{}-space somewhat over time. The prediction used to appropriately position the lunar return distribution in \tdc{}-space is inherently imperfect, and can vary by a few nanoseconds from night to night; this causes the lunar distribution to ``walk'' by as much as $\sim 150$\,\tdc{} bins for collections of runs over multiple nights. The fiducial range photon distribution is better behaved, but can wander by $\sim 7$\,\tdc{} bins over time due to the \fpdSig{} timing dependencies with temperature that cannot completely be eliminated within the temperature-controlled electronics/optics enclosure. We therefore seek to study the correlation between \acsSig{} corrections and the differences in \tdc{} distributions of \fid{} and \lun{} photons\footnote{Note: we did not appreciate this correlation at the time of publication of Ref~\cite{nickThesis}; a different methodology for correcting runs without contemporaneous \acsSig{} data appears there.}.

\subsection{Corrections to ``non-\acsSig{}'' runs}
\label{subsec:corrections_non_acs}

\begin{figure}
        \centering
        \includegraphics[width=0.9\textwidth]{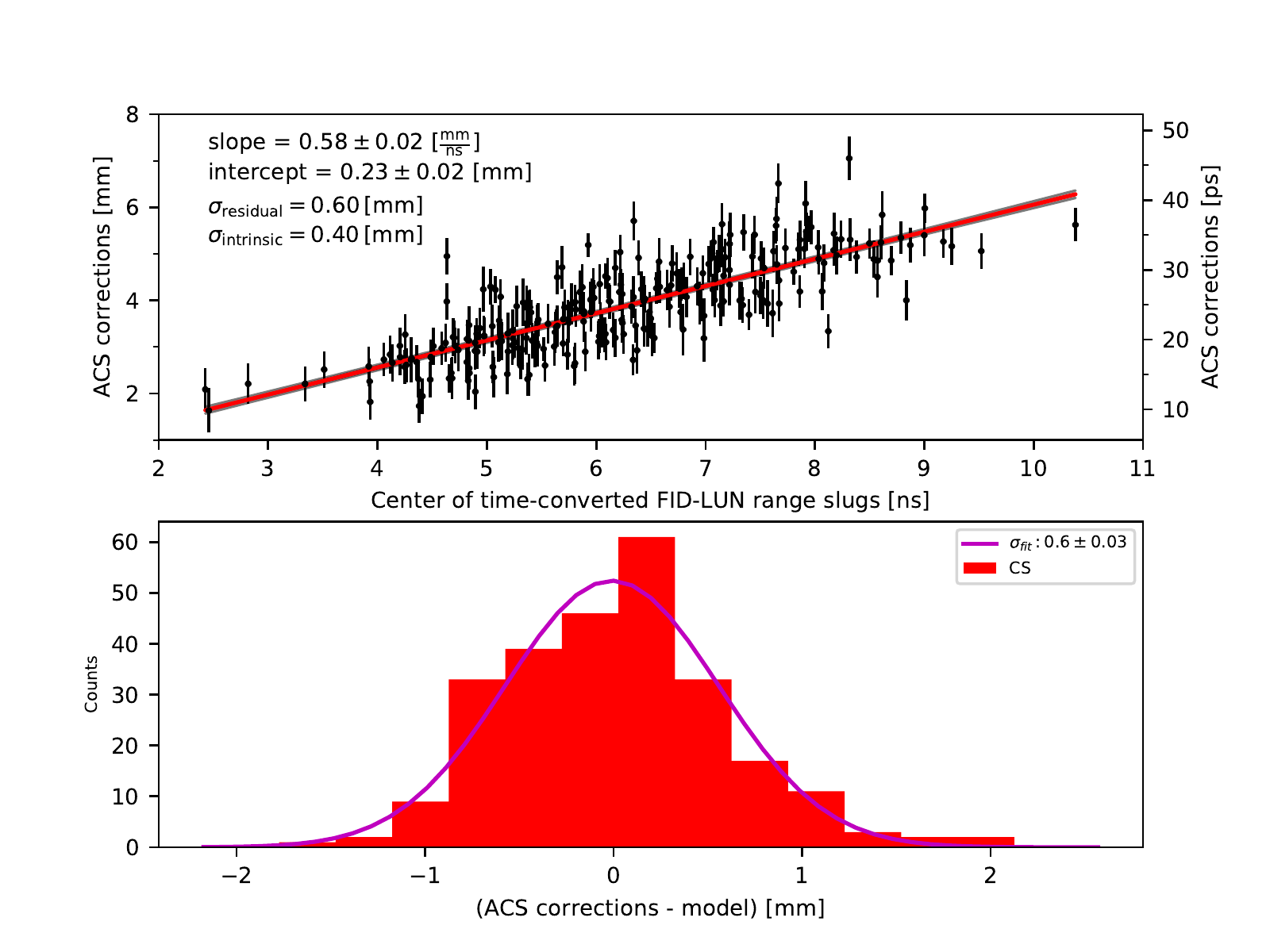}
        \caption{\label{fig:acsCorrs_vs_diffSlugs} Top: \acsSig{} corrections versus differential (fiducial - lunar) position of range photon distributions in \tdc{}-space, converted to time. A fitted line appears in red, with small gray shaded regions indicating the line model uncertainty. Bottom: histogram of the difference between \acsSig{} corrections and the fitted line from the upper subplot.
        }
\end{figure}

During each run's data processing, the positions of range photon distributions (one for each gate) are estimated by calculating the midpoint of the minimum and maximum \tdc{} values of a windowed region over the registered range photons. We can plot \acsSig{} corrections for those runs having contemporaneous \acsSig{} photons against the \fid{} - \lun{} shift and fit a line to it, as shown in the upper subplot of Figure~\ref{fig:acsCorrs_vs_diffSlugs}. The fitted slope comes out to $0.58\,\frac{\mathrm{mm}}{\mathrm{ns}}$, or $\sim 0.4\%$ when put in dimensionless units, which corroborates what was seen in Figures~\ref{fig:toothSep_plot} and~\ref{fig:real_mapping}. It is additionally important to note that when the \fid{} and \lun{} slugs are perfectly overlapped (a condition that can be contrived), APOLLO experiences a minimum of systematic error in the \tdc{}.

The linear fit results can be used to produce a reasonable representation of what the \acsSig{} correction would have been for runs without contemporaneous \acsSig{} data, based on the run's \fid{} - \lun{} shift. An error contribution of $\sqrt{\sigma_{residual}^2 + \sigma_{intrinsic}^2}$ is root-sum-squared into the normal point uncertainty, where $\sigma_{residual}$ is the model residual standard deviation ($0.60\,[\mathrm{mm}]$) which characterizes the scatter of points about the fit line and $\sigma_{intrinsic}$ is a measure of the intrinsic error in \acsSig{} corrections ($0.40\,[\mathrm{mm}]$). The lower subplot of Figure~\ref{fig:acsCorrs_vs_diffSlugs} displays a histogram of \acsSig{} correction residuals, obtained by subtracting the model line from the \acsSig{} corrections, with a gaussian fit function overlaid. The model residual standard deviation appears in the legend of Figure~\ref{fig:acsCorrs_vs_diffSlugs}'s lower subplot; when compared against Figure~\ref{fig:acs_run_correction_compare}, it demonstrates notable improvement in the spread of \acsSig{} corrections.

\subsection{Corrections to ``pre-\acsSig{}'' runs}
\label{subsec:corrections_pre_acs}
We can also extend the usefulness of the non-\acsSig{} correction methodology to historic, or ``pre-\acsSig{}'' APOLLO runs --- those that came before the installation of the \acs{}. The same \tdc{} unit has been used for the entirety of the experiment, so pre-\acsSig{} corrections derived in a similar manner to non-\acsSig{} corrections should still be applicable to old APOLLO data, in principle.

\begin{figure}[tbh]
	\centering
	\includegraphics[width=0.9\textwidth]{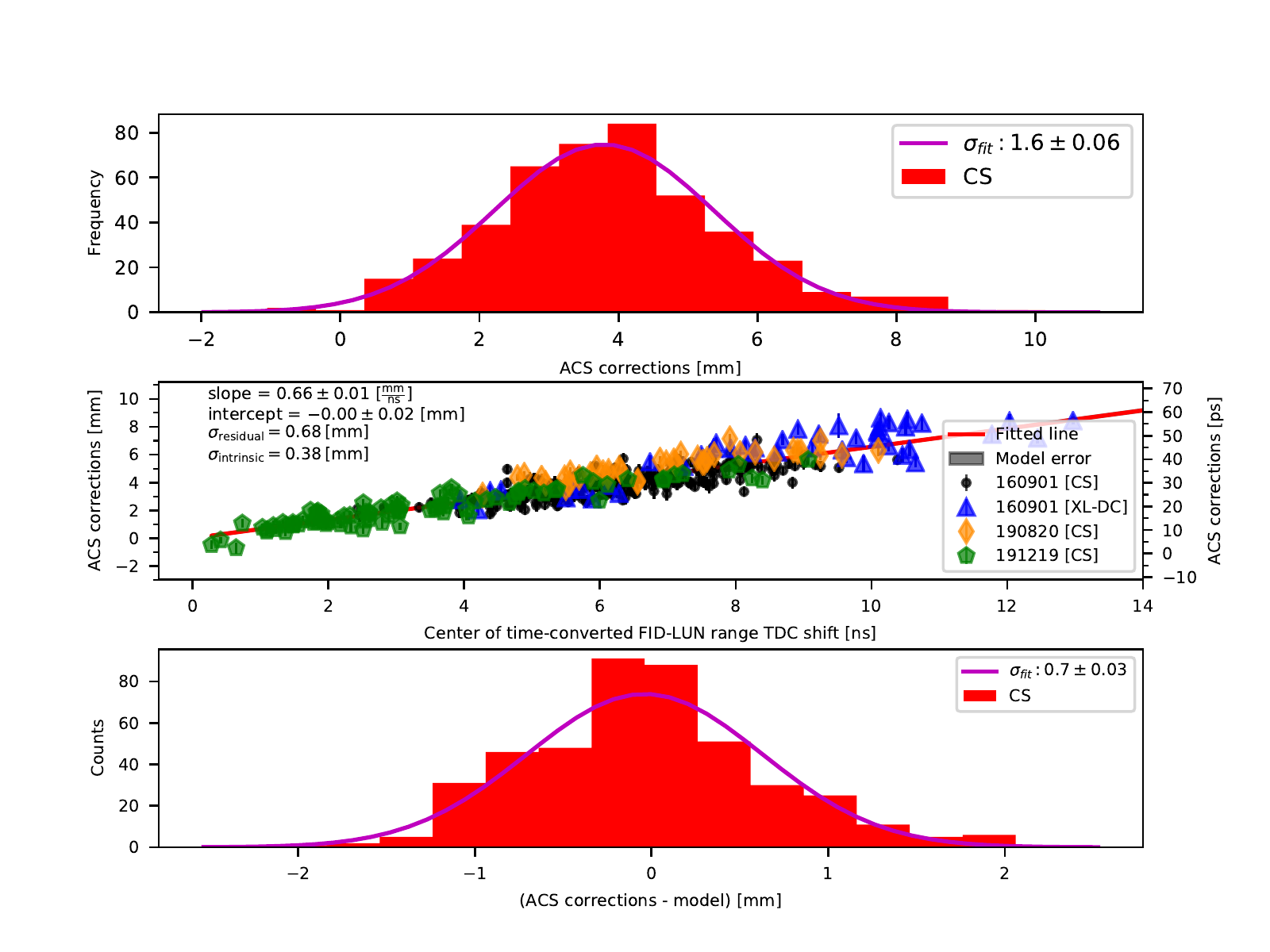}
	\caption{\label{fig:allPeriods_acsCorrs_vs_diffSlugs} A version of the same plot as Figure~\ref{fig:acsCorrs_vs_diffSlugs}, but for all \acsSig{} corrections from 2016 September 12 through 2020 December 27. Note that the black circular markers in the middle subplot are the same data points that appear in Figure~\ref{fig:acsCorrs_vs_diffSlugs}'s upper subplot.}
\end{figure}

The concept remains the same as in Section~\ref{subsec:corrections_non_acs}, except now we work with \acsSig{} correction data from the entire operation of the \acs{} instead of a specific period's data (corresponding to a certain hardware configuration), such as that displayed in Figure~\ref{fig:acsCorrs_vs_diffSlugs}. Figure~\ref{fig:allPeriods_acsCorrs_vs_diffSlugs} shows the result of this concept for all \acs{} runs spanning 12 September 2016 through the end of 2020; the upper subplot parallels Figure~\ref{fig:acs_run_correction_compare}, while the middle and lower subplots parallel Figure~\ref{fig:acsCorrs_vs_diffSlugs}. Four different groups of data appear, corresponding to three different hardware configurations; the earliest \acs{} configuration used either the cesium clock or the XL-DC clock as APOLLO's timebase (distinct from the timebase used to send \acs{} pulses themselves, which is always the cesium clock) at various times, while successive configurations used only the cesium clock as APOLLO's timebase.

Vertical shifts in the groups of data can be seen in the middle subplot of Figure~\ref{fig:allPeriods_acsCorrs_vs_diffSlugs}, corresponding to slightly different systematic errors when the \fid{} and \lun{} signals are perfectly aligned in \tdc{}-space. A single linear fit is performed on the entire collection of data, and we can again estimate what the \acsSig{} correction would have been based on the run's \fid{} - \lun{} shift. The error contribution to the NP is conceptually the same as for non-\acsSig{} runs, but the residual ($0.68\,[\mathrm{mm}]$) and intrinsic ($0.38\,[\mathrm{mm}]$) uncertainties are unique to this fitted line and dataset. The fitted slope of $0.66\,\frac{\mathrm{mm}}{\mathrm{ns}}$, or $\sim 0.4\%$ again demonstrates the newly-discovered systematic undercounting of time by APOLLO, and further substantiates what was previously seen in Figures~\ref{fig:toothSep_plot}, \ref{fig:real_mapping} and \ref{fig:acsCorrs_vs_diffSlugs}.

\subsection{Corrections to non-\acsSig{} and pre-\acsSig{} runs with weak lunar signals}
Corrections to non- and pre-\acsSig{} runs rely on a reasonable estimation of where the fiducial and lunar signals fall in \tdc{}-space. For runs with few lunar return photons, a reliable direct estimation of the lunar signal \tdc{} position may not be possible. However, we may evaluate what a run's lunar signal position was expected to be, based on a few operational variables used in scheduling timing gates, plus a run's offset from the predicted round trip time. Therefore, we wish to study the offset between directly-estimated lunar slug positions and the aforementioned operational variables for runs considered to have trustworthy measures of their lunar slug positions.

\begin{figure}[tbh]
	\centering
	\includegraphics[width=0.9\textwidth]{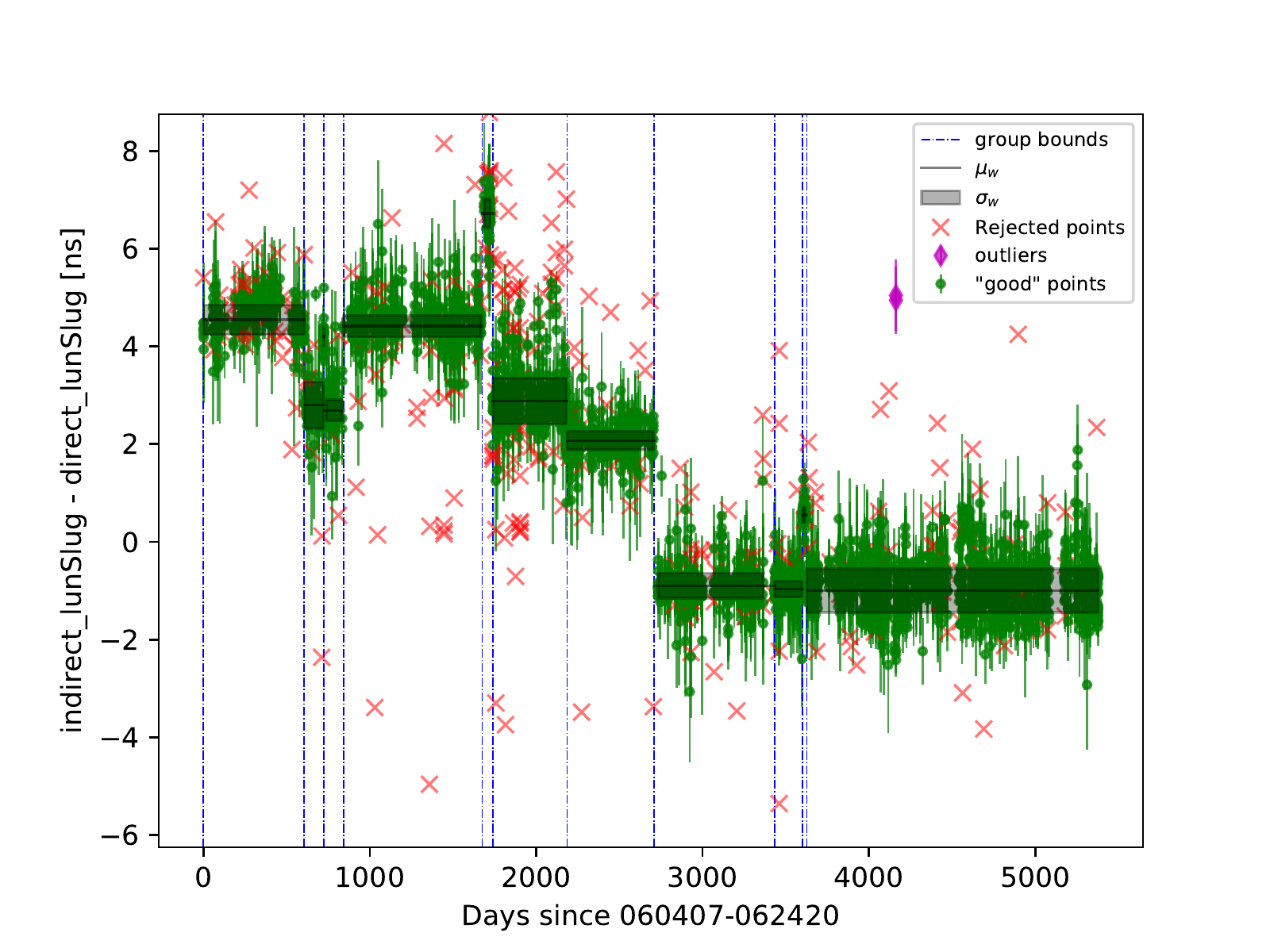}
	\caption{\label{fig:lunSlug_lookup}
	A plot of the offsets between directly-estimated lunar slug positions and indirectly-estimated lunar slug positions, using relevant gate positioning parameters, for runs between 2006 April 7 through 2020 December 27. Vertical shifts in groups of data points, separated by vertical blue dash-dot lines, can clearly be seen. Weighted means for each group appear as solid black lines while shaded gray regions around the solid black lines denote the weighted standard deviation for each group.
	}
\end{figure}

Figure~\ref{fig:lunSlug_lookup} shows the difference between directly and indirectly estimated lunar slug positions for runs from the start of APOLLO through the end of 2020. Green circular markers denote runs whose direct estimate of the lunar slug are considered to be ``good'' based on certain user-defined thresholds of quantities such as estimated number of photons contributing to the lunar signal, the signal-to-noise ratio, how wide the directly-estimated lunar slug is, etc; red $\times$ markers are runs that did not pass these cuts. Magenta diamonds are runs that did pass threshold cuts, but are far enough removed from the green points to not be considered part of them. Data point uncertainties incorporate the estimated error on the direct measure of lunar slug position as well as the prediction offset error (normal point error, pre-\acsSig{}-adjusted).

Vertical shifts in Figure~\ref{fig:lunSlug_lookup} are obvious, and are due to (currently) unidentified/unrecorded changes to the APOLLO system that affected the lunar gate positioning such as temporary changes to the operational software, cable changes, biased channels, etc; these less-than-obvious reasons may be further investigated in the future. However, within each group of points, the mapping between parameters used to indirectly estimate a lunar slug position and the direct measures of lunar slug positions is fairly tight and thus reliable. We know that a lunar slug offset corresponds to \acsSig{} corrections, as demonstrated by Figures~\ref{fig:acsCorrs_vs_diffSlugs} and~\ref{fig:allPeriods_acsCorrs_vs_diffSlugs}. Therefore, the offset and spread of points in Figure~\ref{fig:lunSlug_lookup}, combined with a run's operational variables relevant to gate positioning, are all we need to determine the lunar slug position for lower-signal runs. 

Uncertainties pertinent to indirectly estimating a lunar slug value include the error on the weighted mean and the weighted standard deviation of the appropriate group of points in Figure~\ref{fig:lunSlug_lookup}, as well as the uncertainty of the normal point in question. For example, say we wish to look-up a lunar slug value for the last run performed on 2009 March 16; the relevant group in Figure~\ref{fig:lunSlug_lookup} is the group that includes day 1000. The error on the weighted mean is $\sim 13$\,ps, the weighted standard deviation is $\sim 220$\,ps and the normal point uncertainty is $\sim 38$\,ps to result in a lunar slug uncertainty of $\sqrt{13^2 + 220^2 + 38^2} = \sim 224$\,ps. Next, the looked-up lunar slug value and its uncertainty are used to estimate a differential slug position and a corresponding uncertainty. Finally, an \acsSig{} correction value is estimated as described in Section~\ref{subsec:corrections_non_acs} (or Section~\ref{subsec:corrections_pre_acs} for pre-\acsSig{} runs) and the estimated \acsSig{} uncertainty is RSSd into the normal point uncertainty.

As mentioned earlier, normal points that participate in this estimation of a lunar slug position are chosen based on certain user-defined thresholds of various quantities like the estimated signal to noise ratio. However, even if a normal point is not flagged to use this logic based on threshold conditions, we will check what a looked-up lunar slug uncertainty would be and compare it against the normal point's directly-estimated (the average \tdc{} position of lunar range returns contributing to the signal) lunar slug error. If the looked-up uncertainty is less than the directly-estimated uncertainty, we will utilize the looked-up lunar slug information instead.

\section{Conclusions}
The Earth and Moon play a critical role in studying aspects of our most poorly-understood fundamental force, gravity. LLR has been able to take advantage of these local massive bodies to constrain gravitational parameters that may otherwise be difficult to investigate. Additionally, LLR is sensitive to geodetic studies that involve establishment of the lunar and terrestrial coordinate systems, as well as composition of the lunar interior. APOLLO has contributed high-caliber data to advance these aforementioned pursuits for $15$\,years using a robust method of data reduction, as described in Section~\ref{sec:photon_packaging}.

In 2016, an Absolute Calibration System was added, allowing the independent study of APOLLO timing inaccuracies. The \acs{} is able to run \emph{while collecting} range data, granting the powerful ability to correct timing inaccuracies of the lunar range data itself on a photon-by-photon basis. We are able to extend the \acs{}'s usefulness to those runs lacking in contemporaneous \acsSig{} data by referencing their \fid{} - \lun{} shifts and predicting appropriate timing corrections to apply to their final normal points. Finally, the original purpose of the \acs{} was fulfilled by identifying a significant source of systematic error in the \tdc{} part of APOLLO's timing system; the \tdc{}'s systematic $0.4$\% inaccuracy is multiplied by how imperfectly overlapped the \fid{} and \lun{} signals are to produce a RTT error.

Runs on or after September 12, 2016 with sufficient contemporaneous \acsSig{} data are correctable as described in Section~\ref{sec:rangeCalibrationWithAcs}. Runs on or after September 12, 2016 that lack ample \acsSig{} statistics and runs prior to September 12, 2016 are correctable based on the systematic error \acs{} uncovered, as described in Sections~\ref{subsec:corrections_non_acs} and~\ref{subsec:corrections_pre_acs}, respectively. The data presented in Figure~\ref{fig:acs_run_correction_compare} indicates APOLLO is accurate at the $< 1$\,mm level.
APOLLO normal point statistics are presented in a companion paper,~\cite{APOLLO_NP_paper_2023}. That study indicates that APOLLO operates in the millimetric regime, with a median nightly range precision of $1.0$\,mm since the \acs{} was added in September 2016.

At the beginning of 2021, stewardship of APOLLO was transferred to the National Aeronautics and Space Administration's (NASA) Space Geodesy Project (SGP) at Goddard Space Flight Center (GSFC). Lunar range data taken prior to 2021 will be presented in ``JPL'' (or ``MINI'') format and Consolidated laser-Ranging Data (CRD) format on Zenodo \cite{np_dataset}; CRD-formatted data will additionally appear on the International Laser Ranging Service's (ILRS) global data centers: NASA's Crustal Dynamics Data Information System (CDDIS) and the EUROLAS Data Center (EDC) \cite{ILRS2019}.

\subsection*{Acknowledgments}
We thank Russet J. McMillan for her continued support as APOLLO's operator during observation, and handling of \acs{} functions during ranging. We thank Ed Leon for periodically performing measurements of clock phase at the observatory, and for occasional hardware troubleshooting. We thank the NASA Postdoctoral Program (NPP) for continuing to fund this study after the NASA transition took place; we thank Stephen M. Merkowitz of NASA GSFC's Space Geodesy Project for his aid in facilitating the NPP appointment and support thereafter. We also thank Frank G. Lemoine of NASA GSFC for providing feedback on this publication prior to formal peer review.

This work is based on access to and observations with the Apache Point Observatory 3.5-meter telescope, which is owned and operated by the Astrophysical Research Consortium. This work was jointly funded by the National Science Foundation (PHY-0602507, PHY-1068879, PHY-1404491, PHY-1708215) and the National Aeronautics and Space Administration (NNG04GD48G, NAG81756, NNX12AE96G, NNX15AC51G, 80NSSC18K0482). JBRB acknowledges the support of the NASA Massachusetts Space Grant (NNX16AH49H).

\section*{References}
\bibliographystyle{iopart-num}
\bibliography{apollo}

\end{document}